# Abstract

Recently, we developed a dynamic distributed end-to-end vehicle routing system (E2ECAV) using a network of intelligent intersections and level 5 CAVs (Djavadian & Farooq, 2018). The case study of the downtown Toronto Network showed that E2ECAV has the ability to maximize throughput and reduce travel time up to 40%. However, the efficiency of these new technologies relies on the acceptance of users in adapting to them and their willingness to give control fully or partially to CAVs. In this study a stated preference laboratory experiment is designed employing Virtual Reality Immersive Environment (VIRE) driving simulator to evaluate the behavioral response of drivers to E2ECAV. The aim is to investigate under what conditions drivers are more willing to adapt. The results show that factors such as locus of control, congestion level and ability to multi-task have significant impact.

**Keywords:** virtual immersive reality environment, laboratory experiment, driver behaviour, discrete choice model, connected and autonomous vehicles, distributed routing, intelligent intersections,

# 1.    Introduction

In the past century as countries developed and became more affluent, cars became one of the major essentials for each household. In Ontario, Canada alone, as reported by the Ministry of Transportation of Ontario (MTO) in 2010 there were 8.7 billion registered vehicles (MTO, 2013). This global increase in personal car ownership has given rise to congestion rate, accident rate and greenhouse gas (GHG) emissions level. Between 1990 and 2008, it is reported that road transportation emissions grew by 1.6% per year (MTO, 2013). According to the study conducted by Metrolinx (2008) in 2006, the annual cost of travel delays, increased impact to the environment, and increased chance of vehicles collision to commuters in the Greater Toronto and Hamilton Area was $3.3 billion. In 2006 the cost to the economy in the form of GDP was estimated at $2.7 billion and the estimated costs for 2031 to commuters and economy will balloon to $7.8 billion and $7.2 billion respectively. Therefore, now more than ever, transportation engineers and planners are seeking innovative ways to address these problems.

The advent of communicating intelligent vehicles has opened new doors to manage traffic, reduce congestion and negative environmental impacts of driving while increasing road safety. Transport Canada (2017) recently invested $2.9 million in funding under the program to Advance Connectivity and Automation in the Transportation System in order to help Canadian jurisdictions prepare for connected and autonomous vehicles (CAVs). There are many safety and sustainability benefits to this technology, such as having the ability to avoid collisions without requiring human reaction time or to smooth traffic flow patterns (dubbed "green driving") to reduce excess energy

consumption or emissions. One major application of CAVs is dynamic traffic navigation which provides real time travel time information and status to drivers. Studies have shown that route guidance based on CV can reduce travel time (Yang & Recker, 2006; Park & Lee, 2008; Katan et al., 2012; Claes et al. (2011); Du et al. (2014); Yamashita et al. (2005)). At Laboratory of Innovations in Transportation (LiTrans), Ryerson University, we have developed a novel dynamic distributed End-to-End vehicle routing system using network of intelligent intersections and fully CAVs (E2ECAV) (Djavadian & Farooq, 2018). Applying the new routing to a case study of downtown Toronto, results showed that it is possible to reduce travel time by 40%. However, the efficiency of these new upcoming and disruptive technologies depends on their acceptance by drivers and willingness of drivers to give full or partial control to the CAVs (Hoogendoorn et al, 2014; Kulmala, 2010; De Vos and Hoekstra, 1997). Therefore, there is a strong need to investigate the behavioral response of drivers in CAVs environment.

As stated by Calvert et al, (2017), no empirical studies have been done in an experimental setting to understand this problem. In recent years, online stated preference surveys (SP) have been used to test the factors affecting travelers' acceptance of automated vehicles (e.g. Fagnant et al, (2014); Daziano et al, (2017); Bansal et al, (2016)), for detailed list the interested reader is referred to Becker & Axhausen (2017). However as powerful as SP survey is in analyzing user's evaluation of different alternatives it lacks realism especially when it comes to innovative alternatives (e.g. CAVs, electrical vehicles) where there is no prior reference. In a study conducted by Cherchi and Hensher (2015) and also as pointed out by Professor Elisabeta Cherchi at her Keynote Speech at the International Association of Travel Behaviour Research Conference (2018), there is a need for visual and engaging tools e.g. eyes tracking, virtual reality, and simulators to add value to behavioural relevance and reduce hypothetical bias. As a result, in the past couple of years, the use of virtual reality state preference survey has been gaining interests among researchers since in some extend it allows users to form a visual image of the innovative alternative as opposed just purely mental image. For example, Farooq et al. (2018) in their recent studies used SP virtual reality survey to investigate the interaction of pedestrians with automated vehicles.

The purpose of this study is to address above mentioned key gap. In order to find out what factor affect the adaptation of travelers to the new routing system (E2ECAV) and under what conditions they are more willing to adapt. in this study stated preference laboratory experiments are designed, employing Virtual Immersive Reality Environment (VIRE) driving simulator. We aimed to answer the research question: *Under what conditions the willingness of drivers to adapt to E2ECAV will increase?*

The remainder of this paper is organized as follows. In the background section an overview of our proposed (E2ECAV) is presented. The methodology section presents design of our SP experiments employing VIRE driving simulator. After results and analysis are presented followed by summary and future work directions.

## 2.    Background

As discussed earlier one of the advantages of connected and automated vehicles is the ability to provide dynamic vehicle routing using up-to-date travel time information. However, the efficiency of these routing systems is usually hindered by several factors. One major factor that affects the effectiveness of any routing systems is the compliance of the drivers with the given advice. As shown in previous research (Djavadian et al., 2014; Knoop et al., 2011, Netten et al, 2006), drivers are not always willing to consider and comply with information they receive or share information. Knoop et al. (2011) observed that drivers are much more receptive to mandatory systems as oppose to voluntary systems. Djavadian et al. (2014) reported that drivers are more willing to comply when they are well informed and well rewarded. Mandatory (prescriptive) information high compliance rate whereas voluntarily (descriptive) information low compliance rate unless rewarded. Netten et al., (2006) showed that drivers are more willing to comply when they can acknowledge the gain, for instance, they are more willing to comply with an advice that instructs them to accelerate rather than to an advice that instructs them to slow down. In the case of routing system based on CAVs and vehicle to vehicle communication (V2V), factors such as market penetration rate of CAVs and their communication range limitations can also affect the efficiency of the vehicle routing (Yange & Recker, 2006).

To increase the efficiency of route guidance system using CAVs and address the non-compliance nature of the drivers, market penetration and communication range issues, Djavadian & Farooq (2018) proposed a dynamic route guidance based on a network of intelligent intersections using level 5 CAVs (Gasser & Westhoff, 2012), where full cooperation and coordination can be expected. Under the proposed E2ECAV framework, intersections using infrastructure to infrastructure communication (I2I) are able to obtain real time traffic information from neighboring intersection and adjacent links creating a single integrated and coherent view of the network that is frequently updated. In the proposed system, drivers are no longer decision makers and automated vehicles are guided by intelligent intersections through the network from origin to destination in such a way that the proposed system optimizes network throughput.  Fig. 30.1 presents the schematic view of the proposed E2ECAV.

To test the efficiency of the proposed E2ECAV it was implemented in an in-house agent-based traffic simulation platform, using downtown Toronto network (refer to Fig.30.2) and real travel data from Transportation Tomorrow Survey (DMG). The simulation results showed that E2ECAV is capable of outperforming human driven vehicles and automated vehicles (no communication) by reducing travel time up to 40% in case of re-current congestion and 15% in case of non-recurrent condition (Djavadian & Farooq., 2018). Testing different market penetration levels, the results showed that at high penetration rates (50%-70%) the proposed E2ECAV has significant impact on flow, density and speed, resulting in maximization of capacity. The significant improvement in travel time and capacity was achieved by having single integrated and coherent view of the network that is constantly updated and having 100% compliance of the vehicles.

The effectiveness of the proposed E2ECAV is impacted by the willingness to adapt of drivers and giving control to the new system. In this study we try to answer the question whether drivers will adapt to the proposed system and if they do under what conditions. Moreover, we are interested to know the impact of socio-demographic factor on adaptation of drivers.

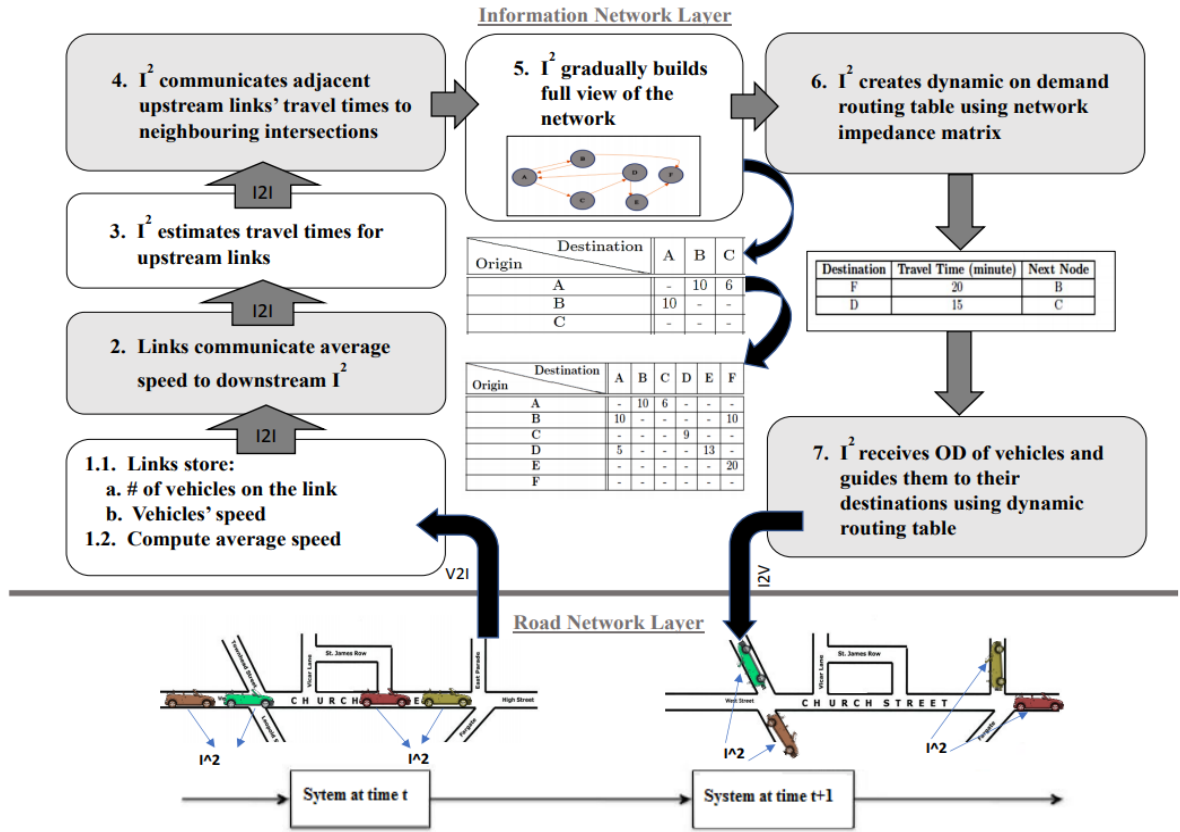

**Fig. 30.1 Flow diagram for E2ECAV routing (Djavadian & Farooq, 2018)**

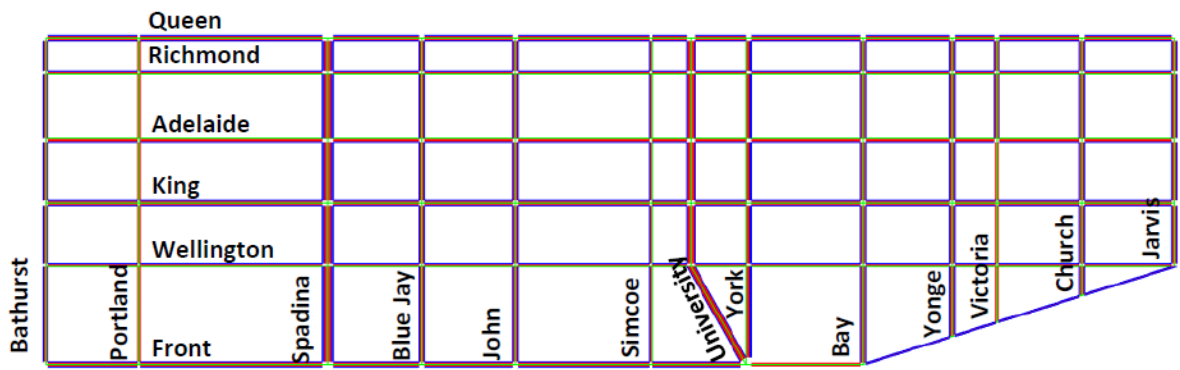

**Fig. 30.2 Simplified downtown Toronto network (Djavadian & Farooq, 2018)**

# 3.    Methodology

In order to answer the posed research question in the introduction section, a stated preference experiment is designed employing VIRE driving simulator (Farooq et al., 2018). This section presents methodology used to design the experiments and model travelers' behaviours.

## 3.1. Objectives

In this study VIRE driving simulator and pre-experiment questionnaire are used to investigate under what conditions drivers are willing to be the passenger of CAV and under what conditions they prefer to drive themselves to their destinations. The aim of VIRE driving simulator is to help us answer the following questions:

- What is the effect of familiarity with road network on acceptance of E2ECAV?
- What is the effect of multi-tasking on acceptance of E2ECAV?
- What is the effect of traffic congestion (low, high) on acceptance of E2ECAV?
- What is the effect of travel time on acceptance of E2ECAV?

Whereas pre-experiment questionnaire is used to answer the following question:
- What are the effects of gender, age, driving experience, risk index, locus of control index on acceptance of E2ECAV?

## 3.2. Dependent & Independent Variables

### 3.2.1 Dependent variable

In this study since the goal is to observe whether users would choose the proposed E2ECAV (Djavadian & Farooq (2018)) over HDV, the independent variable is set to binary: Yes, for choosing E2ECAV and no otherwise.

### 3.2.2. Independent variables

The independent variables used in this study are presented in this Table 30.1. One of the key independent variables used is the Locus of Control index (Rotter, 1996), that groups people in two categories: a) those who believe they have control over events in their lives and b) those who believe there are external forces out of their control affecting their lives. The first group has internal locus of control whereas the second group has external locus of control. The higher the locus of control index means that the person has higher external locus of control. Studies have shown that people with internal locus of control are in general happier and more adaptable (Rotter, 1966). We hypothesize that people with lower locus of control will be more willing to adapt and give full control to autonomous vehicles.

The attributes listed in the right column of Table 30.1, are the ones used to compare the two travel options (HDV & CAV) in terms of congestion level, multi-tasking, and network familiarity. Bansal et al. (2016) also used congestion level and multi-tasking as their variables and showed that these two variables have significant impact on the travellers' willingness to use automated vehicles.  Familiarity with a network has been used by previous studies (e.g. Talaat, 2008) to evaluate route choice behavior of the respondents. Becker & Axhausen (2017) provided a detailed literature review of studies that conduced SP survey on adaptation of AVs and they provided extensive list of variables used.

**Table 30.1 Independent variables**

| User attributes | Travel attributes |
|---|---|
| • Age<br>• Gender<br>• Education<br>• Vision<br>• Employment<br>• Driving experience<br>• Route choice attitude<br>• Risk index<br>• Locus of control index | • Network familiarity<br>• Congestion level<br>• Multi-tasking<br>• Travel time |

## 3.3. Choice Modelling

In this study binary Logit is used to model the behaviour of the travelers using the preliminary results obtained from the pilot study. More advance model will be used in future work. Eqn. 30.1 presents the utility function for each travel option.

$$U_{kn} = \beta_{0,k}^T + \beta_{x,k}^T X_{kn} + \beta_{z,k}^T Z_{kn} + \varepsilon_{kn} \quad (30.1)$$

where:

- $U_{kn}$: expected utility of option $k$ for user $n$;

- $X_{kn}$: set of attributes related to option $k$;

- $Z_{kn}$: set user $n$ attributes, e.g. socio-economic variables;

- $\beta_{x,k}^T, \beta_{z,k}^T$: set of parameters corresponding to the attributes;

- $\varepsilon_{kn}$: unobserved utility modeled as a Gumbel distribution.

. The probability of user choosing E2ECAV is shown by Eqn. 29 2.

$$P(E2ECAV) = \frac{1}{1 + e^{-(\beta_{0,E2ECAV}^T + \beta_{x,E2ECAV}^T X_{E2ECAVn} + \beta_{z,E2ECAV}^T Y_{CAVn})}} \quad (30.2)$$

## 3.4. Experiment Setup

The laboratory experiment is divided into four sessions as shown below:

- **Information session**
  - Questionnaire
- **Learning session**
  - Introduction to VR
  - Familiarization with the driving network
- **Actual experiment**
  - 3 experiment per user
  - 2 scenarios per experiment

- **De-briefing**
    - Feedback from the participants

### 3.4.1. Information session

During the first session participants are provided with information about the experiment and are asked to fill out the pre-experiment questionnaire which consisted of 5 sections as described below.

A. ***Socio-economic/Demographic Attributes***. Collects participant's age, gender, occupation, education level and income level.

B. ***Driving/Navigation Device Experiences***. Collects information regarding real-life driving experiences in terms of years of experience, familiarity with in-car navigation information dissemination, and real-life familiarity with the test network.

C. ***Route Choice Attributes***. Collects information regarding the criteria each participant uses to choose his/her route (e.g.: travel time, distance, mileage, gas).

D. ***Personality Attributes***. Collects participant's attitudes toward adventure and discovery through (Khattak et al., 1995). A risk index is estimated for each subject, based on a scoring system. Alternative answers for each question are given a score from 0 to 4 in an ascending order; starting with 0 for option (i). The risk index, for each subject, is estimated to be the sum of scores of all questions. High risk index indicates a risk-seeking type of personality. Similar test was also used by Talaat (2008).

E. ***Locus of Control***. Collects subject's internal versus external control reinforcement and provides information on personal perception of self-efficacy and control of a situation. The test is developed by Rotter (1966). Scores range from 0 to 13. A low score indicates an internal control while a high score indicates external control.

### 3.4.2. Learning session

The second session is the learning session where participants are asked to drive around the network in order to familiarize themselves with the test network and also learn how to drive in the virtual reality environment.

### 3.4.3. Actual experiment session

The actual experiment is conducted in the VIRE driving simulator. Every participant is asked to try 3 experiments each consisting of 2 scenarios which will be discussed in more details in the next paragraph. At the end of each experiment participants are asked to select which scenario they preferred the most, driving themselves or being driven.

#### *3.4.3.1. Experiment scenarios*

As mentioned earlier, each participant has to go through 3 experiments with 2 scenarios. The scenarios are as follows: HDV and E2ECAV. In the case of both HDV and E2ECAV there are two possible road networks, the familiar network where participant does the learning session and the unfamiliar network different than the learning session network. Further there are two different traffic conditions, low and high. In the case of E2ECAV the users also have the option of multi-tasking and non-multi-tasking. In total there are 8 experiments (two scenarios each) and out of these 8, we randomly assign 3 experiments to each participant in such a way that all experiments are repeated equal number of times. Fig. 30.3 presents the breakdown of possible scenarios for HDV and E2ECAV.

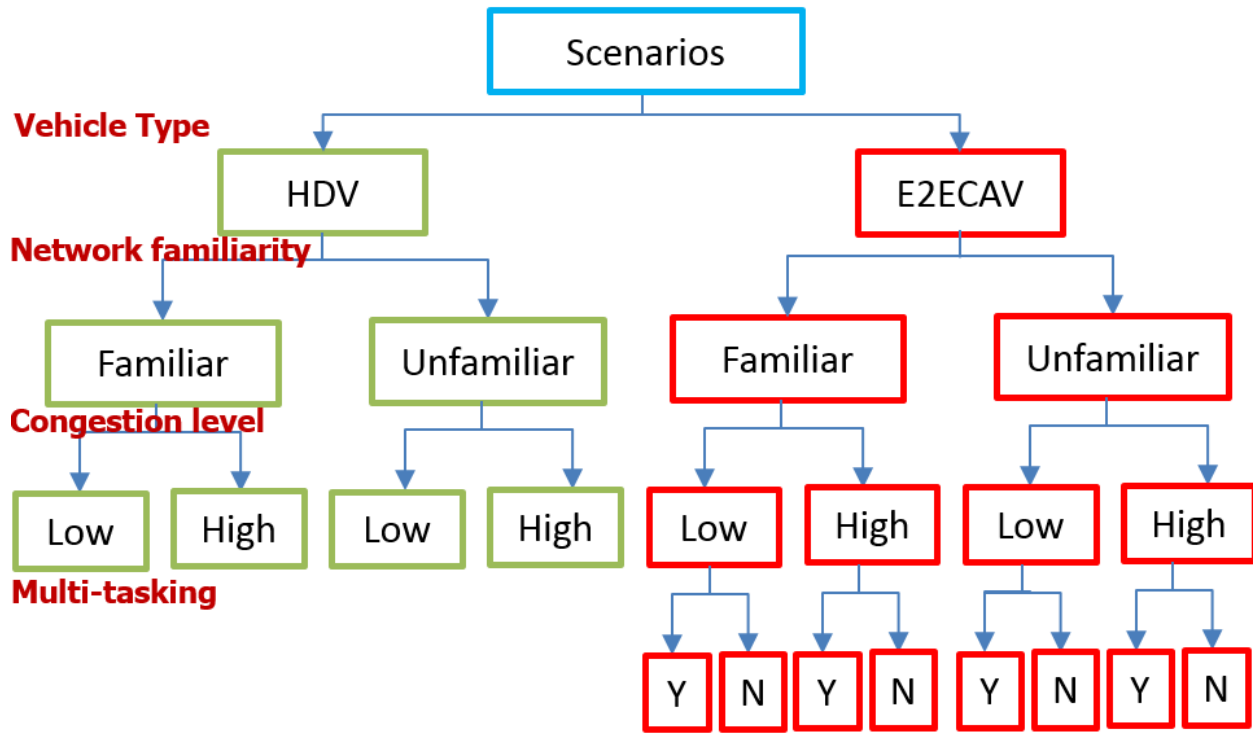

**Fig. 30. 3 HDV & E2ECAV Scenarios**

Before the start of each experiment, participants are assigned an origin-destination pair (in total there are two O-D pairs one for familiar network and one for unfamiliar network). Depending on a scenario they either have to drive themselves or be driven by a CAV. In the case of the HDV scenario they are also given a static map as shown in Fig. 30.4A and B. At the end of each scenario, their travel time is shown on the screen in order for them to compare the two alternatives. The two networks used in this study are both part of downtown Toronto network, which was also used by Djavadian & Farooq (2018) for their case study. The reason for choosing the same network is to utilize the E2ECAV data collected from their agent-based simulation to model the movement of E2ECAV in VIRE simulator.

The two OD pairs used are: intersection of King St. & Simcoe to intersection of Queen St. W. & Bathurst St., and intersection of Wellington St. & University Ave. to Queen St. E. & Jarvis St.

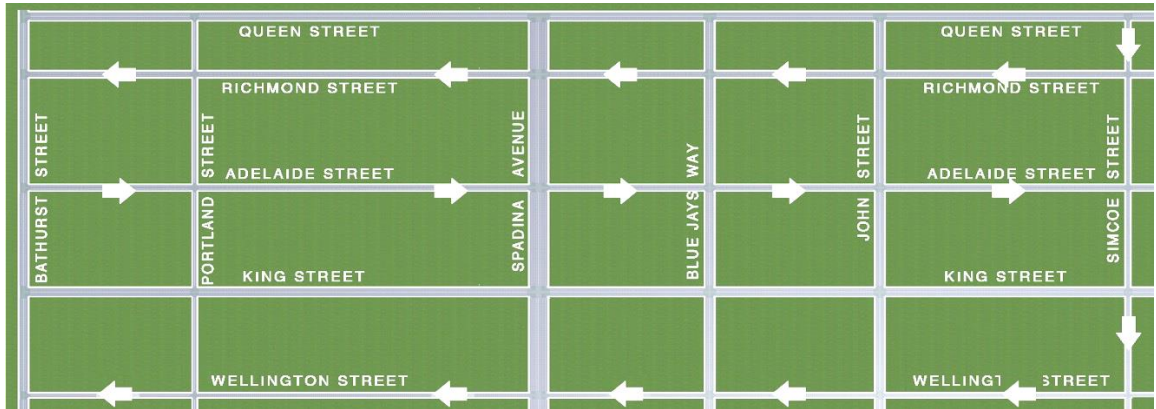

(A)

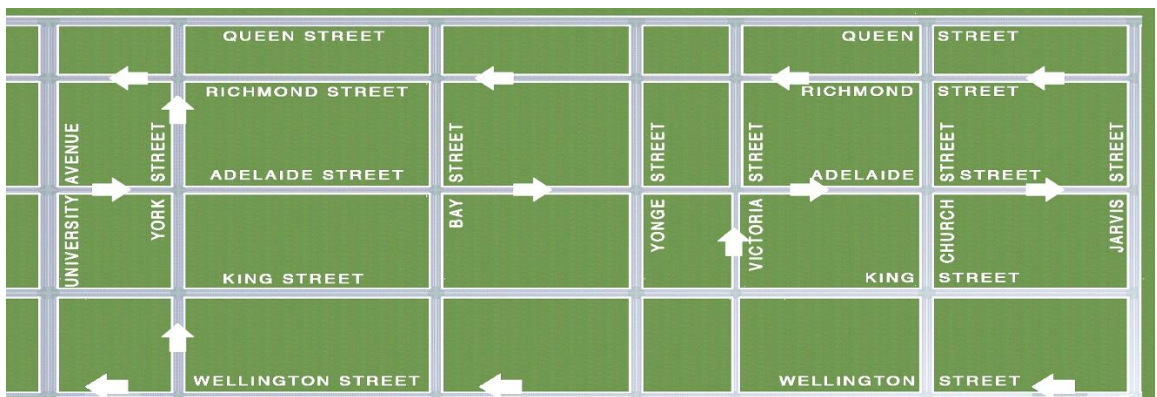

(B)

**Fig. 30. 4 Test networks. (A) Familiar network (from learning session) and (B) Unfamiliar network**

### 3.4.4. De-briefing

At the end of laboratory experiment a short interview is conducted with the participants to receive their feedback regarding the experiment itself and also provide more information regarding the reasons behind selecting their preferred scenarios.

### 3.5. Virtual Immersive Reality Environment

In the open-source gaming engine Unity which VIRE (Farooq et al., 2018) is based on, the downtown Toronto network was coded as a collection of links and nodes mapped such that the movement of CAVs can be modelled seamlessly using the data from Djavadian & Farooq (2018) traffic simulation. For CAV scenarios, the option to multi-task was also introduced such that the user riding in the passenger seat of a CAV can read a virtual newspaper or play a maze game on a virtual phone. VIRE was also modified to include driving hardware such as a steering wheel and acceleration and braking pedals to allow users to drive in the virtual environment. Traffic conditions for the HDV scenarios in which participants drive were simulated by spawning bot-vehicles on the link the user is currently driving on as well as nearby links. The spawning rate is randomized between at arbitrary ranges of time in order to simulate different traffic conditions.

While VIRE is capable of collecting physical data such as the participant's gaze as well as virtual data such as the participants speed, coordinates, the number of collisions or near-collisions, and route used, for the purposes of this case study we only utilized the total trip time.

The following hardware is used:

- Oculus Rift with motion and touch sensors
- Thrustmaster T150 Force Feedback Racing Wheel,
- Intel 7 core processor
- Nvidia GeForce 1080 graphic card

Fig. 30.5 presents the snapshots of our virtual reality setup for the two travel options. Fig. 30. 5A is the snapshot of HDV scenario where as Fig.30. 5B is the snapshot of the CAV scenario.

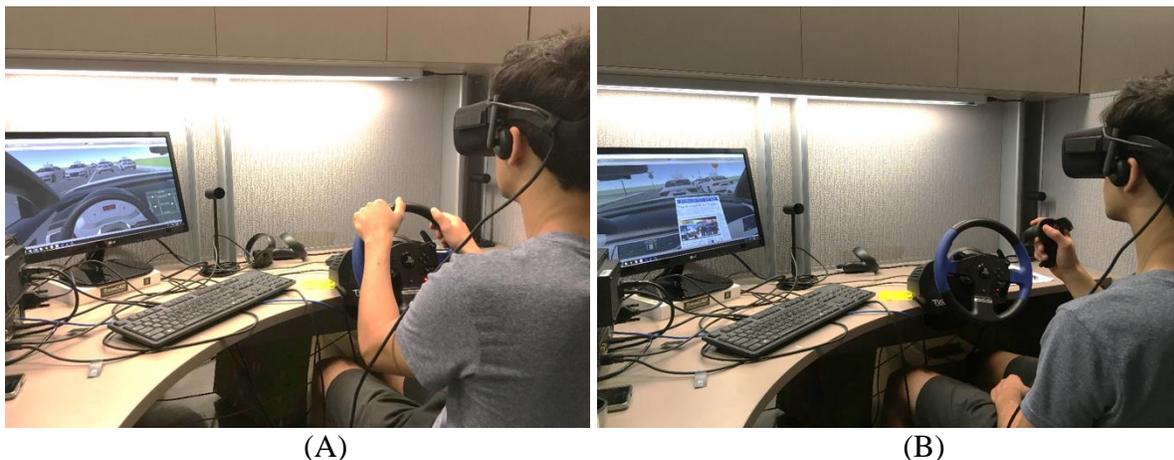

(A)                                   (B)

**Fig. 30.5 Participants in VIRE driving simulator. (A) HDV and (B) E2ECAV with multi-tasking**

# 4. Results & Analysis

This section provides results and analysis of the pilot study tested with graduated students and employees of the Laboratory of Innovations in Transportation (LiTrans) at Ryerson University. The following information is collected from the laboratory experiment:

**Data that collected from the pre-experiment questionnaire**
- Age
- Gender
- Education
- Employment
- Driving experience
- Route choice attitude
- Risk index
- Locus of control index

**Data collected from the VR simulator for each scenario**
- Trip time

**Data collected from the post-experiment questionnaire**
- Participants' choices (HDV or E2ECAV)

## 4.1. Socio-demographic Characteristics of Participants

Table 30.2 provides a brief description of our pilot study participants. In total we had 17 participant who as mentioned above were the graduate students and employees at our department. As can be seen from Table 30.2, the collected sample is homogenous in terms of participants' socio-demographic characteristics and driving experience.

**Table 30.2 Socio-demographic characteristics of the participants**

| | Number of Participants, by Age (years) | | | |
|---|---|---|---|---|
| **Characteristics** | **18-24** | **25-29** | **30-39** | **Total** |
| **All participants** | 9 | 2 | 6 | 17 |
| **Gender** <br> Female <br> Male | 3 <br> 6 | 0 <br> 2 | 3 <br> 3 | 6  (35%) <br> 11 (65%) |
| **Occupations** <br> Student <br> Employee | 9 <br> 0 | 2 <br> 0 | 3 <br> 3 | 14 (82%) <br> 3  (18%) |
| **Education** <br> Bachelor <br> Masters <br> Doctorate | 6 <br> 3 <br> 0 | 0 <br> 1 <br> 1 | 2 <br> 0 <br> 4 | 8 (47%) <br> 4 (24%) <br> 5 (29%) |
| **Driving experience (years)** <br> Not at all <br> <2 <br> 2-5 <br> 5-10 <br> > 10 | 1 <br> 2 <br> 5 <br> 1 <br> 0 | 0 <br> 0 <br> 0 <br> 2 <br> 0 | 0 <br> 0 <br> 1 <br> 0 <br> 5 | 1 (6%) <br> 2 (12%) <br> 6 (35%) <br> 3 1(8%) <br> 5 (29%) |

## 4.2. VIRE Driving Simulator Results

Table 30.3 provides a number of times the proposed E2ECAV was chosen by the participants over HDV under different experiment setting. In total there were 43 observations, and as can be seen from Table 30.3 out of that 19 times participants chose E2ECAV over HDV, which is 43% of the time. From Table 30.3 it can be observed that participants chose E2ECAV mostly when they were asked to drive on the un-familiar network and when they had the option of multi-tasking. It may be surprising to see that they chose E2ECAV under low traffic conditions, because intuitively we would think it should be the opposite. For example, a study conducted by Bansal et al., (2016) showed that travelers are more willing to let autonomous vehicle drive them on highways and in highly congested traffic condition. The reason for this discrepancy is the way robot vehicles and CAVs are modeled in the VIRE at the moment. Since CAVs are based on traffic simulation results they portrayed congestion more accurately whereas robot vehicles are spawned at arbitrary rate, this on occasion caused the HDV network to be less congested than CAV network. This issue will be addressed in the future studies.

**Table 30.3 Choice of E2ECAV over HDV by participants**

| Experiment # | HDV | | E2ECAV | | E2ECAV selected |
|:---:|:---:|:---:|:---:|:---:|:---:|
| | Familiar network/ Un-familiar network | Traffic Congestion | Multi-tasking/ No-multi-tasking | Traffic Congestion | |
| 1 | Fam | Low | Multi | Low | 3 |
| 2 | Fam | Low | Not-multi | Low | 2 |
| 3 | Fam | High | Multi | High | 3 |
| 4 | Fam | High | Not-multi | High | 0 |
| 5 | unFam | Low | Multi | Low | 5 |
| 6 | unFam | Low | Not-multi | Low | 3 |
| 7 | unFam | High | Multi | High | 3 |
| 8 | unFam | High | Not-multi | High | 0 |
| | | | | | 19 |

## 4.3. Binary Logit Model Results

The initial results of binary Logit model estimation based on the data gathered from the pilot study are presented in Table 30.4. The parameter values are not finalized, this is just to provide us with an idea of what variables had the most impact on the choices of participants when it came to the preference of E2ECAV over HDV. As can be seen, the three main factors where locus of control, congestion level and multi-tasking.

In terms of locus of control as discussed earlier people with lower locus of control index having internal locus of control and more adaptable, this explains the negative sign for locus of control parameter. The higher the index the lower the adaptability. With respect to multi-tasking if it was available the variable took the value of 1 and otherwise 0. It was shown in Table 30.3 that most participants chose E2ECAV when multi-tasking was available this explains the positive sign of multi-tasking variable. Similar results obtained from a SP survey conducted by Bansal et al (2016) were 75% of the respondents concluded that they would like to do multi-tasking such as texting, talking with friends or look out of the window during their ride. In terms of congestion, the value was 1 if it was low congestion and else 0. Based on the preliminary results participants preferred E2ECAV over HDV because it chose the less congested route than they used.

**Table 30.4 Preliminary Binary Logit Model estimation results**

| Variables | Estimate | t-stat |
|---|---|---|
| Locus of control (0-12) | -0.16 | -2.16 |
| Congestion level (1,0) | 1.40 | 2.10 |
| Multi-tasking (1,0) | 1.10 | 1.71 |

| | |
|---|---|
| # of Observations | 43 |
| # of parameters tested | 3 |
| Likelihood ratio test | 9.348 |
| Adjusted Rho-square | 0.056 |

# 5. Conclusion

In this study we designed and conducted a stated preference experiment employing virtual immersive reality environment simulator to evaluate the willingness of drivers to adapt to CAVs and answer the question under what conditions users' willingness increases.

In this study binary Logit model was used to model the behaviour of the travelers and their adaptation to CAVs using the preliminary results obtained from the pilot study conducted with graduate and employees at the Laboratory of Innovations in Transportation, Ryerson University. The results showed that factors such as locus of control, multi-tasking and traffic condition have significant impact on the willingness to adapt of the drivers. Based on preliminarily results, drivers were more willing to adapt and give control to CAV when multi-tasking option was available to them. Furthermore, those with lower locus of control index were also more willing to adapt than those with higher locus of control index, this is because those with lower index tend to believe that control comes from within rather than externally as such they adapt faster to new situations. In fact, those who chose HDV over CAV they mentioned one major factor that affected their choice was that they felt more in control when they were driving themselves as opposed to being passenger of CAV. Based on the results obtained drivers were more willing to choose CAV when traffic condition was low this is because they felt CAV taking faster and less congested route. We expected that more participant choose CAV under congested traffic conditions, however in this study our experiment set up and calibration may have affected the choices of the participants. In future studies we will look into replicating CAV traffic condition for HDV as well.

There are several directions that can be taken in future studies. First and foremost, the laboratory experiment will be conducted with larger and more heterogeneous sample size. Next the data from Djavadian & Farooq (2018) agent-based simulation study will be used to model the traffic condition for HDV scenarios to be closer to the same traffic condition experienced under E2ECAV. We had few participants that they experienced motion sickness in different levels, and we are hoping that by calibrating our VR and including driving rig we be able to provide more comfortable experience for our participants.

Future study will also investigate joint Diffusion and Latent class with Mixed Logit model for choice model estimation. Although Logit model is powerful in evaluating demand for different alternatives, it is based on SP market share and does not consider evolution of adaptation over time especially in the case of new alternatives. Diffusion model on the other hand is powerful in forecasting adaption over long run for new innovations however it uses basic demand model. Studies have shown that joint Diffusion and Logit model is more powerful that each of them individually. For example, Jensen et al. (2017) used joint discrete choice model and diffuse model to predict the potential for electric vehicles.

## Acknowledgement

This research is funded by NSERC Canada Research Chair on Disruptive Transportation Technologies and Services, Ontario Early Researcher Award, and Ryerson University. We would also like to thank Yoser Hamid the undergraduate summer intern at LiTrans for providing us with support during this project.